\documentclass[twocolumn,showpacs]{revtex4}

\usepackage{graphicx}
\usepackage{dcolumn}
\usepackage{amsmath}

\makeatletter
\def\btt#1{\texttt{\@backslashchar#1}}
\DeclareRobustCommand\bblash{\btt{\@backslashchar}}
\makeatother

\begin{document}

\title[]{Gravitational collapse of Type II fluid in higher dimensional
 space-times}

\author{S. G. Ghosh}
\thanks{To whom all correspondence should be addressed.}
\affiliation{%
 Department of Mathematics, Science College, Congress Nagar,
 Nagpur-440 012, INDIA}%

\author{Naresh Dadhich}
\email{nkd@iucaa.ernet.in} \affiliation{Inter-University Center
for Astronomy and Astrophysics, \\
 Post Bag 4, Ganeshkhind, Pune - 411 007, INDIA}%

\date{\today}

\begin{abstract}
We find the general solution of the Einstein equation for
spherically symmetric collapse of Type II fluid (null strange
quark fluid) in
higher dimensions.
It turns out that the nakedness and curvature strength of the
shell focusing
singularities carry over to higher dimensions. However, there
is
shrinkage of
the initial data space for a naked singularity of the Vaidya
collapse due to the
presence of strange quark matter.
\end{abstract}


\maketitle

\section{Introduction}
The gravitational collapse of sufficiently massive star, under
fairly
general conditions, will result into a singularity is the fact
established
by the singularity theorems of Hawking and Ellis \cite{he}.
However these
theorems do not indicate whether the singularity will be hidden
behind the
event horizon or whether it will be visible to an outside
observer. The singularity
will not be visible is indicated in the relativistic literature
by the phrase
{\it  cosmic censorship conjecture} (CCC). It will not be
visible to an observer
outside the event horizon is the weak form while it will not be
visible to any
observer, even the one who is sitting on the collapsing star,
is
the strong form
of the conjecture.  This conjecture was articulated by Penrose
\cite{rp}
(see \cite{r1} for reviews).  The weak form essentially states
that gravitational
collapse from a regular initial data never creates the
space-time singularity
visible to distant observers, i.e., any singularity that forms
must be hidden within
a black hole. For the strong form on the other hand, nothing is
supposed to emanate
out of the singularity and hence it is not visible to anyone
even if one is
infinitesimally close to it.

There is no general theory of nature and visibility of
singularities. There do exist a number of exact solutions of the
Einstein equation which admit depending upon the initial data
black holes (BH) or naked singularities (NS) (see, e.g., Joshi
\cite{r1}). In particular the Vaidya solution \cite{pc} is
extensively used to show that the end state of collapse for a
regular initial data results into a naked singularity. In the
absence of a general result, the study of various examples of
collapse becomes pertinent to examine the validity of the
phenomenon of occurrence of naked singularities. The two most
investigated cases are that of inhomogeneous dust and the Vaidya
null fluid. A more general matter distribution should be
considered. Undoubtedly, as singularity is approached, matter
would be in highly dense and exotic state. The strange quark
matter is the densest state of matter known which is produced in
quark-hadron phase transition in the early Universe or at ultra
high energy neutron stars converting into strange quark matter
stars \cite{ew}. It would therefore be interesting to study the
collapse of the mixture of the Vaidya fluid with the strange
matter (SQM), which is a Type II fluid \cite{he}. This has been
studied in the usual 4-dimensional space-time \cite{gd1} and here
we wish to study it in higher dimensions. The presence of SQM
tends to shrink the initial data window leading to NS. As is
expected, the increase in dimensions would also tend to enhance
the shrinkage because of gravity getting stronger in higher
dimensions \cite{gd}.

Recent developments in string theory indicate that gravity may be
truly higher-dimensional(HD) interaction, becoming effectively
$4D$ at lower energies. The main question is how to bring this
theoretical development on the anvil of observations? One of the
ways is to study its effects in  astrophysical settings. That is
where the phenomenon of gravitational collapse in HD attains
relevance. In this study we would be finding the effects of HD as
well as SQM on the end product of collapse of null SQM.  For this,
we first obtain the exact solution describing collapse of null SQM
fluid sphere, and then investigate their effect in the context of
CCC.

In section II, we obtain the generalized Vaidya solution
for null quark fluid  in higher
dimensional spherically symmetric space-time.
For definiteness we shall refer this metric as HD generalized
Vaidya metric.
We adapt this solution
to study end state of collapse in section III .  Finally we
shall conclude
with a discussion on the effect of SQM on HD Vaidya collapse.

\section{Type II fluid collapse in higher
dimensions}
Let us begin with the general $(n + 2)$-dimensional spherically
symmetric space-time, in advanced Eddington time coordinate
$v$,
described by the metric \cite{pd,jr}:

\begin{equation}
ds^2 = - e^{\psi(v,r)} dv \;\left[ f(v,r) e^{\psi(v,r)} dv + 2dr
\right] + r^2 d {\Omega}_n^2 \label{eq:me1}
\end{equation}
where $0 \leq r \leq \infty$ is the proper radial coordinate,
$-\infty \leq v \leq \infty$ is an avanced time coordiante, and
\begin{eqnarray}
&& d\Omega^2 = \sum_{i=1}^{n} \left[ \prod_{j=1}^{i-1} \sin^2
\theta_j \right] d \theta_i^2 = d \theta_1^2+ sin^2 \theta_1 d
\theta_2^2 + \nonumber \\
&& sin^2 \theta_1 sin^2 \theta_2 d\theta_3^2+\;.\; .\;
.\;  \nonumber \\
&& + sin^2 \theta_1 sin^2 \theta_2 .\; .\; . sin^2 \theta_{n-1}
d\theta_n^2 \label{eq:ns}
\end{eqnarray}
is the metric on $n$-sphere in spherical polar coordinates.  It
proves useful to introduce a local mass function $m(v,r)$ defined
by $f=1 - 2m/r$ \cite{bi}. Here $m(v,r)$ is an arbitrary function
of advanced time $v$ and radial coordinate $r$. When $m = m(v)$,
it is the Vaidya solution in higher dimensions \cite{pd,gd}. The
usual Vaidya solution in 4-space-time follows for  $m = m(v)$ and
$n=2$.

We wish to find the general solution of the Einstein equation for
the matter field given by Eq.(\ref{eq:me}) for the metric
(\ref{eq:me1}), which contains two arbitrary functions. Hence it
is general in the retarded coordinates used. It is the field
equation $G^0_1 = 0$ that leads as in the $4D$ case as well as in
the present HD case to $ e^{\psi(v,r)} = g(v)$ \cite{gd1}. This
could be absorbed by writing $d \tilde{v} = g(v) dv$.  Hence,
without loss of generality, the metric (2) takes the form,
\begin{equation}
ds^2 = - \left[1 - \frac{2 m(v,r)}{(n-1)r^{n-1}}\right] d v^2 + 2
d v d r + r^2 d {\Omega}_n^2 \label{eq:me}
\end{equation}

The energy-momentum tensor for the Type II (null SQM) fluid is
given by \cite{ww,vh},
\begin{equation}
T_{ab} = \mu l_al_b +(\rho + p)(l_a\eta_b + l_b\eta_a) + p
g_{ab}
\label{eq:emt}
\end{equation}
\begin{eqnarray}
l_{a} = \delta_a^0, \: n_{a} = \frac{1}{2}
\left[ 1  - \frac{2 m(v,r)}{r} \right ] \delta_{a}^0 -
\delta_a^1
\nonumber \\
l^a = \delta_1^a,  \: n^a = - \delta_{0}^{a} -  \frac{1}{2}
\left[ 1  - \frac{2 m(v,r)}{r} \right ] \delta_{1}^a \nonumber
\\
l_{a}l^{a} = n_{a}n^{a} = 0 \; ~l_a n^a = -1.
\end{eqnarray}
Here $\rho,~ p$ are the strange quark matter energy density and
thermodynamic
pressure while $\mu$ is the energy density of the Vaidya null
radiation.
The null vector $l_a$ is a double null eigenvector of $T_{ab}$.
Physically occurring distribution is null radiation flowing in
the radial
direction corresponding to $\rho = p = 0$, the Vaidya
space-time
of
radiating star. When $\mu = 0$, $T_{ab}$ reduces to degenerate
Type I
fluid \cite{pd}, and it represents string dust for $\mu = 0 =
p$.
The energy condition for such a distribution are as follows
\cite{he}: \\
\noindent
(i) the weak and strong energy conditions,
\begin{equation}
\mu > 0, ~\rho \geq 0, ~p \geq 0 \label{eq:ecw}
\end{equation}
(ii) the dominant energy condition,
\begin{equation}
\mu > 0 ~\rho \geq p \geq 0 .\label{eq:ecd}
\end{equation}
By choosing properly the mass function $m(v,r)$, it is possible
to satisfy the energy
conditions.  In the particular case of $m = m(v)$, the energy
condition reduces to
$\mu \geq 0$.

The field equations are \cite{pd},
\begin{subequations}
\label{fe1}
\begin{eqnarray}
&& 8\pi\mu   = \frac{n \dot m}{(n-1)r^n},   \label{equationa}
\\
&& 8\pi \rho = \frac{n  m ^{\prime } }{(n-1)r^n},
\label{equationb} \\
&& 8 \pi p   =  -\frac{ m ^{\prime \prime}}{(n-1)r^{n-1}}.
\label{equationc}
\end{eqnarray}
\end{subequations}
Here and in
what follows dash and dot denote derivative respectively
$\partial / \partial r$ and $\partial / \partial v$.
The part $\mu l_al_b$ of $T_{ab}$ in (\ref{eq:emt}) is the
component of
matter field that moves along the null hypersurface $v =
const.$. In
particular when $p = \rho = 0$ we have the Vaidya solution in
higher
dimensions. Thus the distribution in (\ref{eq:emt}) represents
Vaidya
radiating star in Type II fluid universe in higher dimensions.
Note that
when $\mu = 0$, it is static perfect fluid space-time, which
will yield
regular space-time only when $m\propto r^{n+1}$ which is the de
Sitter
space. As in the $4D$ case, the equation of state for the zero
temperature quark
matter is taken \cite{hc}:
\begin{equation}
p = \frac{1}{k}  (\rho - 4 B) \label{eq:eos}
\end{equation}
where $k>0$ is a constant. Inserting the above equation in the
field
equations, we find that
\begin{equation}
m'' = - \frac{nm'}{k r} + \frac{32 \pi B}{k} r^{n-1}.
\end{equation}
If we choose the function $m(v,r)$ such that
\begin{equation}
m(v,r) = m_{0}(v,r) + \frac{\Lambda}{(n+1)} r^{n+1}
\label{eq:mf},
\end{equation}
where  $m_{0}(v,r)$ is an unknown function, then we find that
\begin{equation}
m_{0}''= - \frac{n}{k} m_{0}',  \hspace{.2in}
B = \frac{32 \pi \Lambda}{n(k+1)}\label{eq:de}.
\end{equation}
It admits the general solution
\begin{equation}
m_{0} = S(v)r^{-n/k+1} + M(v). \label{eq:sde}
\end{equation}
Now, one can easily
calculate $m$, $\mu$, $\rho$ and $p$, which are
\begin{eqnarray}
&& m = M(v)+S(v) r^{-n/k+1} + \frac{\Lambda}{n+1} r^{n+1}  \\
&& \mu = \frac{n}{8\pi (n-1) r^n}
\left[ \dot{M}(v)+\dot{S}(v)r^{-n/k+1}\right]      \\
&& \rho = \frac{n}{8\pi (n-1) r^n} \left[\frac{k-n}{k}
S(v) r^{-n/k}+ \Lambda r^n \right] \\
&& p =  \frac{n}{8\pi (n-1) r^n} \bigg[\frac{(k-n)}{k^2}
S(v) r^{-n/k+1} -
\Lambda r^n \bigg].
\end{eqnarray}
This is the general solution of collapsing null strange quark
fluid
in higher dimensions.  The metric explicitly reads as
\begin{eqnarray}
&& ds^2= - \bigg[1 - \frac{2 M(v)}{(n-1)r^{(n-1)}}-
\frac{2S(v)}{(n-1)r^{[(n-2)k+n]/k}} - \nonumber  \\
&& \frac{\Lambda}{n^2-1} r^{2} \bigg] dv^2 + 2 dv dr + r^2 d
\Omega^2.
\end{eqnarray}
Clearly, all the energy conditions would be satisfied for $n
\geq 2$ because
it would ensure $\rho \geq0$ and $p \geq0 $, while $\mu \geq0$
would be taken
care of when we choose the mass functions for both null
radiation and SQM.
The case, $S(v)=0$, corresponds to the HD Vaidya de Sitter
solution \cite{pd,gd},
whereas the Vaidya solution is obtained by taking  $S(v) = B =
0$.  The solutions in
Refs. \cite{gd1,hc} can be recovered by setting $n=2$.
In the next section, we shall adapt the
solution obtained to study the end state of the collapse for
$k=3$
in Eq. (\ref{eq:eos}), for which the matter is quark plasma.

\section{Existence of a naked singularity}
The physical scenario here is that of a radial influx of null
fluid
in an initially empty region of the higher dimensional
non-flat but  empty  space.
The first shell
arrives at $r=0$ at time $v=0$ and the final at $v=T$.
A central singularity of growing mass is
developed at $r=0$.
For $ v < 0$ we have $m(v)\;=\;0$, i.e., the higher
dimensional
de Sitter like space,
and for $ v > T$,
$\dot{m}(v)\;=\;0$, $m(v)\;$ is positive
definite.  The metric for $v=0$ to $v=T$ is the higher
dimensional
generalized Vaidya metric, and for $v>T$ we have the
higher dimensional Schwarzschild solution. For the HD Vaidya
region, we choose,
\begin{subequations}
\label{FC}
\begin{eqnarray}
&& 2 M(v) =  \alpha(n-1) v^{n-1} \;  (\alpha > 0),
\label{FCa}  \\
&& 2 S(v) =  \beta (n-1) v^{[(n-2)k+n]/k} \; (\beta > 0),  \;
\label{FCb}
\end{eqnarray}
\end{subequations}

Radial null geodesics of the metric (\ref{eq:me}) must satisfy
\begin{equation}
\frac{dr}{dv} =  \frac{1}{2}  \bigg[1 - \frac{2
M(v)}{(n-1)r^{(n-1)}} -
\frac{2S(v)}{(n-1)r^{[(n-2)k+n]/k}} -
\frac{\Lambda}{n^2-1} r^2 \bigg].   \label{eq:de1}
\end{equation}
Clearly the above differential equation has a singularity at
$r=0$, $v=0$.
We note that a singularity forms when a
shell hits the origin at $r= 0$, where the density diverges.
If the singularity is at least locally naked (for brevity we
have
addressed it as naked in this paper), there must be a light ray
coming out from it. The critical ray is the Cauchy horizon
defined
as the first outgoing radial null geodesic emerging from the
singularity. Therefore, by investigating the behavior of radial
null geodesics near the singularity, one can find out if
outgoing
null geodesics meet the singularity in the past.
In order to determine the nature of the limiting value of $y$
at
$r=0$, $v=0$ on a singular geodesic, we let $ y_{0} = \lim_{r
\rightarrow 0 \; v\rightarrow 0} y = \lim_{r\rightarrow 0 \;
v\rightarrow 0} {v}/{r}$. Using Eq.~(\ref{eq:de1}) and
L'H\^{o}pital's rule we get
\begin{eqnarray}
&& y_{0} =  \lim_{r\rightarrow 0 \; v\rightarrow 0} y =
\lim_{r\rightarrow 0 \; v\rightarrow 0} \frac{v}{r} =
\lim_{r\rightarrow 0 \; v\rightarrow 0} \frac{dv}{dr}=  \nonumber \\
&& \lim_{r\rightarrow 0 \; v\rightarrow 0} \frac{2}{1 - \alpha
y^{n-1} - \beta y^{n(k+1)/k} - \frac{\Lambda}{n^2-1} r^2}
\label{eq:lm2}
\end{eqnarray}
which can be rearranged as
\begin{equation}
\beta y_{0}^{[(n-2)k+n]/k+1} + \alpha y_{0}^n - y_{0} +2 = 0
\label{eq:ae}
\end{equation}
This algebraic equation is the key equation which governs the
behavior of the tangent vector near the  singular point.

If the singularity is naked, Eq.~(\ref{eq:ae}) must have one or
more positive  roots $y_0$, i.e., at least one outgoing
geodesic
that will terminate in the past at the singularity. Hence in
the
absence of positive  roots, the collapse will always lead to a
black hole. Thus, the occurrence of positive roots implies that
the strong CCC is violated, though  not necessarily the weak
one.

\begin{widetext}

\begin{table}
\caption{Variation of critical parameter $\alpha_c$ and $y_0$ with
$D$ in HD Vaidya and HD null SQM Collapse} \label{table1}
\begin{ruledtabular}
\begin{tabular}{c|c|c|c|c}
$D=n+2$ & \multicolumn {2} {c|} {HD Vaidya Collapse} &
\multicolumn {2} {c|}
{HD null SQM Collapse ($\beta=0.001$, $k=3$)}    \\
\colrule

&$\alpha_C^v$ & Double Roots($y_0$)& $\alpha_C^s$ & Double
Roots($y_0)$  \\
\colrule
4 & 0.125      & 4            & 0.12437012771 &4.00339\\
5 & 0.037037   & 3            & 0.036037037037&3.0\\
6 & 0.0131836  & 2.66667      & 0.01179754129 &2.65859\\
7 & 0.00512    & 2.5          & 0.0032850449803&2.47188\\
\end{tabular}
\end{ruledtabular}
\end{table}

\end{widetext}

The limit $\beta \rightarrow 0$ corresponds to HD Vaidya
collapse \cite{gd},
and in that case, Eq.~(\ref{eq:ae}) admits positive roots for
$\alpha \leq
\alpha_C^v$. Hence singularities are naked for $\alpha \in (0,
\alpha_C^v]$, and black holes form otherwise. Thus $\alpha =
\alpha _ C^v$ is
the critical value at which the
transition occurs, the end state of collapse switches from
naked singularities to black holes. This is in agreement with
earlier work (see, e.g., \cite{gd}).

We want to investigate the changes, at least qualitatively,  in
the above picture of  HD Vaidya collapse in the presence of a
SQM. It is verified
numerically that  as $n$  increases, the threshold value
$\beta_T$ decreases
(see Table~\ref{table2}).
For $\beta > \beta_T$, it is BH for all $\alpha$.  Further, the
critical parameter
$\alpha_C^s$ for HD SQM is smaller than corresponding  critical
parameter
$\alpha_C^v$ for HD Vaidya
, i.e, the naked singularity
spectrum of the collapsing HD Vaidya region gets  covered
and hence the initial data space $(0, \alpha_C^v]$ for
a naked singularity in HD Vaidya collapse contracts due to the
presence of
SQM. Thus,
at least at a qualitative level, the presence of SQM
facilitates the formation of black holes. However, it is
interesting to see that for each $\beta$, there exists a
$\alpha_C^s$
such that singularities
are always naked for all $\alpha \in (0, \alpha_C^s]$. That is
for each
$\beta$ there exists a non zero measure set of $\alpha$ values
giving
rise to NS and consequently violating CCC (see Tables
~\ref{table1}, \ref{table2}).
\begin{center}
\begin{table}
\caption{Variation of $\beta_T$ with $D$. For $\beta >
\beta_T$,
the end state of collapse is a black hole for all $\alpha$
($\alpha_T^C = 10^{-6}$)
and $k=3$}
\label{table2}
\begin{ruledtabular}
\begin{tabular}{c|c|c}
$D=n+2$ & Critical Value $\beta_T$ & Equal Roots  $y_0$ \\
\colrule
4 & 0.20519540362596 & 4.99999 \\
5 & 0.037036037037   & 3.0  \\
6 & 0.009547734436   & 2.6  \\
7 & 0.002807259781   & 2.42857  \\
8 & 0.00088477111    & 2.33334  \\
\end{tabular}
\end{ruledtabular}
\end{table}
\end{center}

\subsection{Other Cases}
\paragraph{} Case $k=n$. \\
This corresponds to
\begin{equation}
\mu = \frac{n}{8 \pi (n-1) r^n}
\left[ \dot{M}(v) + \dot{S}(v) \right],  \hspace{.2in}
\rho = - p = \frac{n \Lambda}{8 \pi (n-1)}
\end{equation}
and the algebraic equation takes the form
\begin{equation}
(\alpha +  \beta) y_0^n - y_0 + 2 = 0       \label{eq:ae1}
\end{equation}
The metric in this case takes the form of the
HD Vaidya - de Sitter metric, i.e., the HD collapse of null
fluid in an expanding de Sitter back ground
where $\Lambda$ is generated by the bag constant.  The singularity
is naked for
\[
(\alpha +  \beta) \leq \lambda_C =
\frac{1}{n} \left(\frac{n-1}{2n} \right)^{n-1}
\]
The equal roots at $\lambda_C$ is $y_o = 2n/(n-1)$.
It is remarkable to note that both critical parameters and
tangents to outgoing geodesics are dependent on dimension of
space-time.
Further, we note that $y_0$ is bounded below by value 2, $y_0
\rightarrow 2$
as $\lambda \rightarrow 0$ or $D \rightarrow \infty$.
\paragraph{} Case $k\rightarrow\infty$. \\
We have
\begin{eqnarray}
&& \mu =  \frac{n}{16 \pi (n-1) r^2} [\alpha y^{n-2}+\beta
y^{n-3}],
\hspace{.2in} p=0, \nonumber \\
&& \rho = \frac{n \beta}{16 \pi r^2} y^{n-2} \label{eq:ni}
\end{eqnarray}
and the algebraic equation becomes:
\[
\alpha y_0^n + \beta y_0^{n-1} - y_0 + 2 = 0,
\]
For $n=2$, this would admit a positive root for $\alpha \leq 1/8
(\beta-1)^2$,
giving the
range for naked singularity as obtained in \cite{jdj}. This is
simply the
null fluid collapse in the background of constant potential which
is
characterized by $T^0_0 = T^1_1 = const./r^2$, as is the case in
Eq.(\ref{eq:ni}) above. The condition for naked singularity
modifies to
$\beta^2 < 4 \alpha + 108 \alpha^2 + 36 \alpha \beta + \beta^2$
when $n=3$.

\paragraph{} Other $k$ \\
We also note that as $k$ increases, so does the threshold value
$\beta_T$.
However, to conserve the space we shall not present the details.

\section{Discussion}
Motivated by the development of superstring and other field
theories, there
is considerable interest in the models with extra dimensions
from viewpoint of
both cosmology \cite{rs} and gravitational collapse \cite{hd}.
In this paper,
We have shown that the 4D spherically symmetric solution
describing
null SQM fluid go over to $(n+2)$-dimensional spherically
symmetric
solution and essentially retaining its physical behavior.
The solution obtained includes many previously known solutions
to
Einstein equations, such as, HD Vaidya de Sitter solution.
We have also used this solution to study the end state of
collapsing
star and
showed that there exists a regular initial data which leads to
naked
singularity. Higher dimensions tend to shrink the naked
singularity
initial data space
alternatively enlargement of black hole initial data space,
of corresponding $4D$ collapse.
We have also studied the effect of SQM on
Vaidya collapse (both in $4D$ and $HD$). The  naked singularity
spectrum of Vaidya collapse
shrinks further with introduction of SQM and thus atleast
qualitatively
SQM favors black hole in comparison to naked singularities.
It is also interesting to see that there exists threshold value
$\beta_T$,
which decreases with increase in $n$, such that for  $\beta >
\beta_T$, the
end state of the collapse is always black hole for all
$\alpha$.
However the threshold touches zero only as
$n\rightarrow\infty$. That means for finite dimensions, there
would always be a window howsoever narrow in the initial data
leading to NS. However for sufficiently large $n$, it would
practically be black hole and CCH would be obeyed. This case
includes several previous cases of
spherical collapse in $4D$ and $HD$.

\acknowledgements One of the authors (SGG) would like to thank
IUCAA, Pune for hospitality while this work was done.

\end{document}